# Simultaneous muon and reference hadron measurements in the Compressed Baryonic Matter experiment at FAIR

Anna Senger[1] for the CBM Collaboration

[1] Facility for Antiproton and Ion Research, Darmstadt, Germany

**Abstract**

The mission of the Compressed Baryonic Matter (CBM) experiment at the future Facility for Antiproton and Ion Research (FAIR) in Darmstadt is to explore the QCD phase diagram at high net baryon densities likely to exist in the core of neutron stars. The CBM detector system is designed to perform multi-differential measurements of hadrons and leptons in central gold-gold collisions at beam energies between 2 and 11 A GeV with unprecedented precision and statistics. In order to reduce the systematic errors of the lepton measurements, which generally suffer from a large combinatorial background, both electrons and muons will be measured with the same acceptance. Up to now, no di-muon measurements have been performed in heavy-ion collisions at beam energies below 158A GeV. The main device for electron identification, a Ring Imaging Cherenkov (RICH) detector, can be replaced by a setup comprising hadron absorbers and tracking detectors for muon measurements. In order to obtain a complete picture of the reaction, it is important to measure simultaneously leptons and hadrons. This requirement is fulfilled for the RICH, which has a low material budget, and only little affects the trajectories of hadrons on their way to the Time-of-Flight (TOF) detector. In contrast, the simultaneous measurement of muons and hadrons within the same experimental acceptance poses a substantial challenge. This article reviews the simulated performance of the CBM experiment for muon identification, together with the possibility of simultaneous hadron measurements.

## 1. Introduction

The mission of the Compressed Baryonic Matter (CBM) experiment at the future Facility for Antiproton and Ion Research (FAIR) [1] is to explore the QCD phase diagram at high net baryon densities expected to exist in the core of neutron stars. The CBM research program includes the study of the nuclear matter equation-of-state, and the search for new phases of QCD matter. The most promising observables are multi-strange (anti-) hyperons, lepton pairs, charmed particles, collective flow of identified particles, fluctuations of conserved quantities, and hyper-nuclei. Figure 1 depicts a sketch of the detector configuration of the CBM experiment.

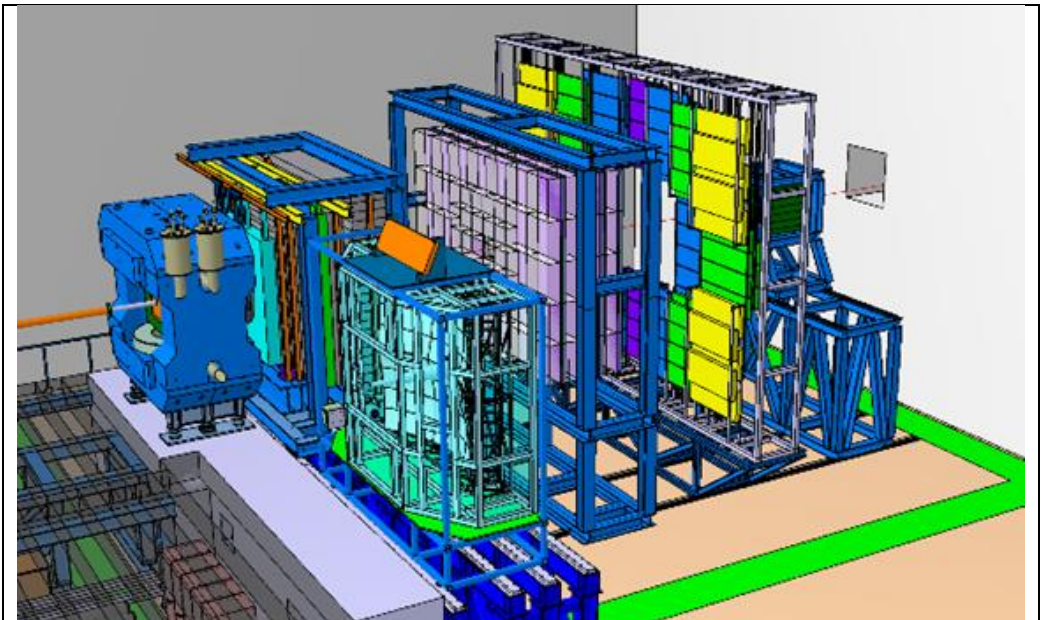

Fig. 1: The CBM experimental setup at FAIR (see text).

The CBM setup consists of two silicon tracking devices located in the magnetic field of a large aperture superconducting dipole magnet. The first tracker is the Micro-Vertex Detector (MVD), which comprises four layers composed of Monolithic Active Pixel Sensors (MAPS). The second device is a Silicon Tracking System (STS), consisting of eight layers covered with about 900 double-sided micro-strip sensors. The total material budget of the STS is equivalent to radiation lengths, which range from 2.4 % and 10.3 %, depending on the vertical distance from the beam axis. The resulting momentum resolution is in the order of 1.5%. Downstream the magnet the lepton detectors are located. The CBM experiment comprises both electron and muon detector systems, in order reduce the systematic uncertainties of the di-lepton measurements. Di-electrons will be identified with a Ring Imaging Cherenkov (RICH) detector, in combination with a Transition Radiation Detector (TRD). Pions with momenta up to about 7 GeV/c will be suppressed mainly by the RICH, whereas the TRD provides pion suppression above 7 GeV/c. The muons will be identified by a Muon Chamber (MuCh) system, which consists of several hadron absorber layers with tracking chambers in between. In figure 1, the MuCh is in measuring position, whereas the RICH is in parking position. These two detectors will be operated alternatively, in contrast to the TRD, which serves also as a tracking detector behind the last muon absorber, as illustrated in figure 1. For hadron identification, a Time-of-Flight (TOF) detector consisting of Multi-gap Resistive Plate Chambers is located behind the TRD. The orientation of the reaction plane will be determined with the help of the Projectile Spectator Detector (PSD), which is located behind the TOF wall. The PSD is a hadron calorimeter consisting of 44 modules, each composed of 60 lead/scintillator sandwiches with 4 mm thick scintillator tiles and 16 mm lead plates. A more detailed description of the CBM detector system, including the novel data-readout and acquisition system, can be found in [2].

The measurement of lepton pairs in heavy-ion collisions provides the unique opportunity to determine the average temperature of the fireball produced in the reaction. The invariant mass spectra of di-leptons is not modified by the collective expansion of the fireball, and, hence, its spectral slope reflects directly the fireball temperature, if di-lepton contributions from other sources are carefully subtracted. The HADES collaboration has demonstrated, that the di-electron invariant mass spectrum - measured in Au+Au collisions at 1.25A GeV - can be corrected by known contributions from vector meson decays. The resulting exponential spectral slope reflects a temperature of about 72 MeV, which represents the average temperature of the fireball, integrated over the collision history [3].

In conclusion, the measurement of the di-lepton excitation function at FAIR energies opens the possibility to determine the caloric curve of hot and dense QCD matter, which would be a direct experimental signature for a first order phase transition. The CBM experiment at FAIR will extend the measurement of di-lepton spectra to higher invariant masses up to about 2.5 GeV/c$^2$, where the spectrum is not contaminated by di-leptons from vector meson decays, which will provide an even more direct access to the average fireball temperature. The measurement of high invariant dimuon masses covers also the region of charmonium, which can be identified for the highest FAIR beam energies using the maximum number of hadron absorbers.

In addition to the dileptons, it is important to measure simultaneously the hadrons for reference. For electron measurements this is easily possible, as the hadrons reach almost undisturbed the TOF due to the low material budget of the RICH and TRD detector. However, in the case of muon measurements, most of the hadrons will be absorbed. Short-lived hadrons like hyperons still can be identified by the STS, without identification of their decay products by the time-of-flight measurement. Moreover, some hadrons still pass the hadron absorbers and reach the TOF detector. This article describes the performance of simultaneous measurements of muons and reference hadrons in the CBM experiment.

## 2. The Muon Detection System of CBM

The measurement of muons in heavy-ion collisions at energies of 2 - 11A GeV is a particular challenge because of the soft muon momenta and the high particle multiplicities. Muon pairs have been measured by the NA60 experiment at the CERN-SPS in Indium-Indium collisions at 158A GeV, which is so far the lowest heavy-ion beam energy where muons could be identified [4]. In order to measure soft muons, e.g. from vector meson decays, in a large combinatorial background, the CBM experiment will feature a variable number of hadron absorbers with tracking detector triplets in-between. The number of installed absorbers and detector stations will depend on the beam energy. Three optional configurations of absorbers and tracking stations are shown in figure 2, which depicts, in addition to the magnet and the MuCh, also the TRD and the TOF detectors. For beam energies up to 4A GeV, a system as illustrated in the left panel of figure 2 will be implemented: 60 cm carbon (density 2.26 g/cm$^3$) + three tracking stations + 20 cm iron + three tracking stations + 20 cm iron + TRD. This MuCh system will allow to study the dimuon invariant mass region up to about 1 GeV/c$^2$. In order to study same invariant mass region at beam energies above 4A GeV, one more iron absorber of 30 cm thickness will be installed, together with two more tracking detector triplets, as shown in the center panel of figure 2. The investigation of intermediate and high dimuon invariant masses ($m_{inv} = 1 - 3$ GeV/c$^2$) at beam energies above 4A GeV requires an additional iron absorber of 1 m thickness, as shown in the right panel of figure 2. The total absorber thickness of this configuration is equivalent to 11.6 hadronic interaction lengths.

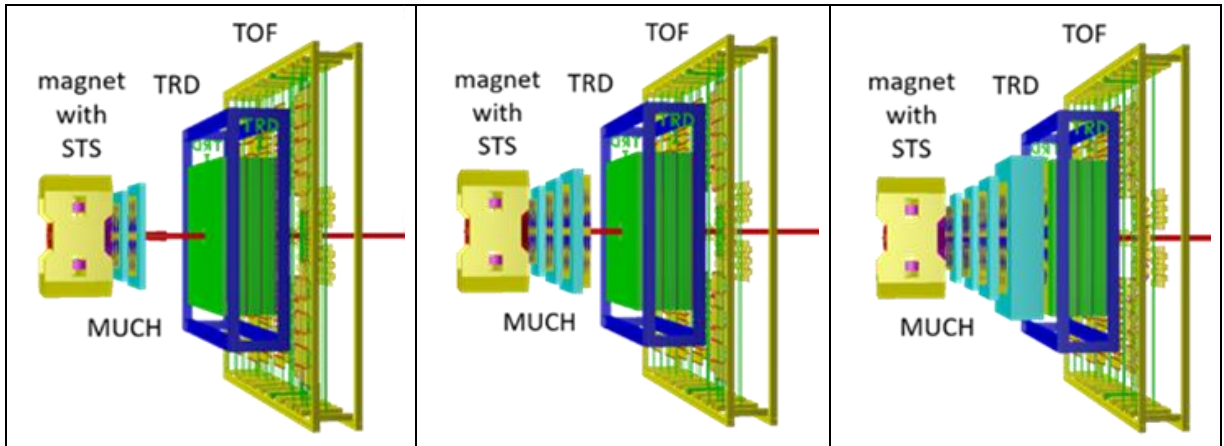

Fig.2: The muon detection located system downstream the magnet, optimized for different beam energies and dimuon invariant mass regions. In addition to the MuCh also TRD and TOF detectors are shown. Left panel: 3 absorber layers + 2 tracking stations for beam energies below 4A GeV, and $m_{inv} \leq$ 1 MeV/c$^2$. Center panel: 4 absorber layers + 4 tracking stations for beam energies above 4A GeV, and $m_{inv} \leq 1$ MeV/c$^2$. Right panel: 5 absorber layers + 4 tracking stations for beam energy above 4A GeV, and $m_{inv} \geq 1$ MeV/c$^2$ (see text).

The technology of the gaseous muon tracking detectors is matched to the hit density and rate. Behind the first and second hadron absorber, where the particle density at small emission angles reaches values of up to 500 kHz/cm$^2$, we will install Gas Electron Multiplier (GEM) Detectors. Prototype GEM detectors with single-mask foils have been successfully tested with particle beams at CERN. The GEM stations are of circular shape with the beam pipe in the center. The first GEM station consists of 16 trapezoidal detector modules with a length 80 cm each, whereas the second, larger station comprises 20 modules with a length of 97 cm. The readout consists of pads with progressively increasing pad sizes from ≈ 3 mm to ≈ 17 mm. The production of the modules has already started. Further downstream at stations 3 and 4, where the peak hit density is reduced to about 15 kHz/cm$^2$ and 4 kHz/cm$^2$, respectively, the use of single-gap RPC detectors with electrodes from low-resistivity Bakelite is under investigation. Additional information on the MuCh hardware developments are presented in [5]. The MuCh project is realized by 13 Indian institutions under the leadership of VECC and Bose Institute in Kolkata, and by PNPI Gatchina.

### 3.     Muon identification

The muon simulations are performed for central Au+Au collisions, which is the most challenging case because of the high track density up to 700 charged particles per event. The hadronic background including muons from weak pion and kaon decays is calculated with the UrQMD event generator [6], whereas the embedded signal distributions from vector mesons are calculated with the PLUTO generator [7], based on the particle multiplicities taken from the PHSD microscopic off-shell transport approach [8]. The contribution of dimuon radiation from a possible Quark-Gluon Plasma has been calculated with a coarse-grain approach [9]. The GEANT3 transport code is used to propagate the created particles through the setup, taking into account the proper matter distribution. The hits in the STS, the MuCh, the TRD and the TOf detector are used for track reconstruction based on a Cellular Automaton algorithm, which is part of the CBM reconstruction software package. The particle tracks passing all absorber stations are considered as muon candidates. Moreover, the information from the TOF detector is used to reduce the background from hadrons which punch though all detector systems. This procedure is illustrated in figure 3, which depicts the squared mass of the particles reaching the TOF detector versus their momentum, for muons from omega meson decays (left panel), and for hadrons (right panel). The mass-square distribution of the dimuons from ω mesons is used to optimize a polynomial fit, which then is used to reduce the background from kaons and protons in the simulation, as shown in the right panel of figure 3.

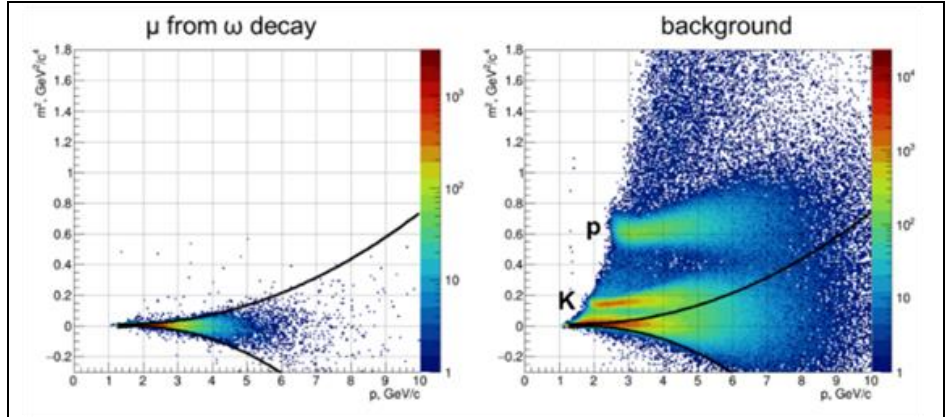

Fig.3: Mass squared as a function of momentum for muons from omega meson decays (left), and for background particles (right).

The resulting dimuon invariant mass spectra simulated for central Au+Au collisions based on the MuCh configurations shown in figure 2 are depicted in figure 4. For the magnetic field the nominal value of 1 T was chosen, and a 250 μm thick Au target was assumed. The left panel of figure 4 depicts the result for a beam momentum of 4A GeV/c, and a MuCh configuration of 3 absorber layers and 2 tracking stations. The spectrum in the center panel of figure 4 corresponds to a beam momentum of 8A GeV/c, calculated for 4 absorber layers and 4 tracking stations. In the right panel of figure 4, the spectrum was calculated again for a beam momentum of 8A GeV/c, but based on 5 absorber layers and 4 tracking stations. The colored histograms indicate the simulation input of the di-muon signals from η, ω, φ, and ρ meson decays including their Dalitz decays, and the thermal radiation from the Quark-Gluon Plasma (QGP). The black histograms represent the result of the reconstruction, including all signal pairs together with the combinatorial background. A particular promising result is the excellent signal-to-background ratio in the order of 10 of the QGP contribution for the MuCh configuration with 5 absorbers as shown in the right panel of figure 4, demonstrating that the CBM detector system is very well suited to measure the radiation from the hot and dense fireball in heavy-ion collisions at SIS100 energies.

The acceptance and reconstruction efficiency for muons from ω decays has been simulated for central Au+Au collisions at a beam momentum of 8A GeV/c. The results are shown in figure 5 for a configuration with 4 (red symbols) and 5 absorbers (blue symbols). The upper three panels depict the values for muons from ω decays as function of the muon momentum, whereas the lower three panels refer to the acceptance and reconstruction efficiency for ω mesons as function of the ω momentum. As mentioned above, the setup with 5 absorbers is used for the measurement of invariant dimuon masses above 1 GeV/c2 including the J/ψ meson, and is shown here for comparison only. The left vertical column depicts the 4π acceptance for muons (upper panel) and ω mesons (lower panel), the central column illustrates the reconstruction efficiency for accepted muons and ω mesons, and in the right column the efficiency with respect to 4π is shown.

The left upper panel of figure 5 illustrates, that for the configuration with 4 hadron absorbers (red symbols) the acceptance for muons from ω meson decays reaches values above 60% at low momenta, and then decreases towards high momenta. This is due to fact, that muons are boosted in beam direction, resulting in high-momentum muons emitted in small angles, and disappearing into the beam hole of the first layers of the silicon stations. The reconstruction efficiency for accepted muons is about 50% at low momenta, and decreases slowly to 20% at high momenta (upper center panel, red symbols). Finally, the muon reconstruction efficiency with respect to 4π decreases from about 40% at low momenta to a few percent at high momenta (upper right panel, red symbols).

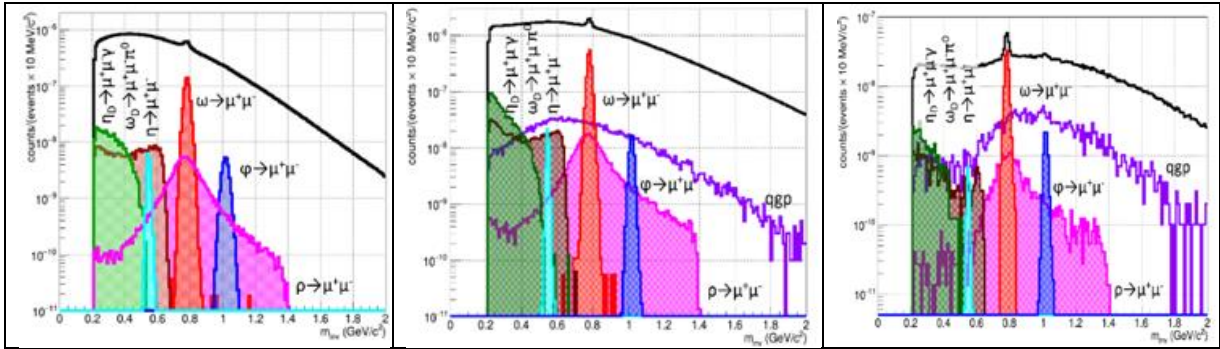

Fig. 4: Dimuon invariant mass spectra simulated and reconstructed for central Au+Au collisions for beam momenta of 4A GeV/c (left panel) and for 8A GeV/c (center and right panel). The black histogram depicts the sum of signal pairs and combinatorial background. The signal contributions are based on MC information. The used MuCh configurations correspond to the ones shown in figure 2 (see text).

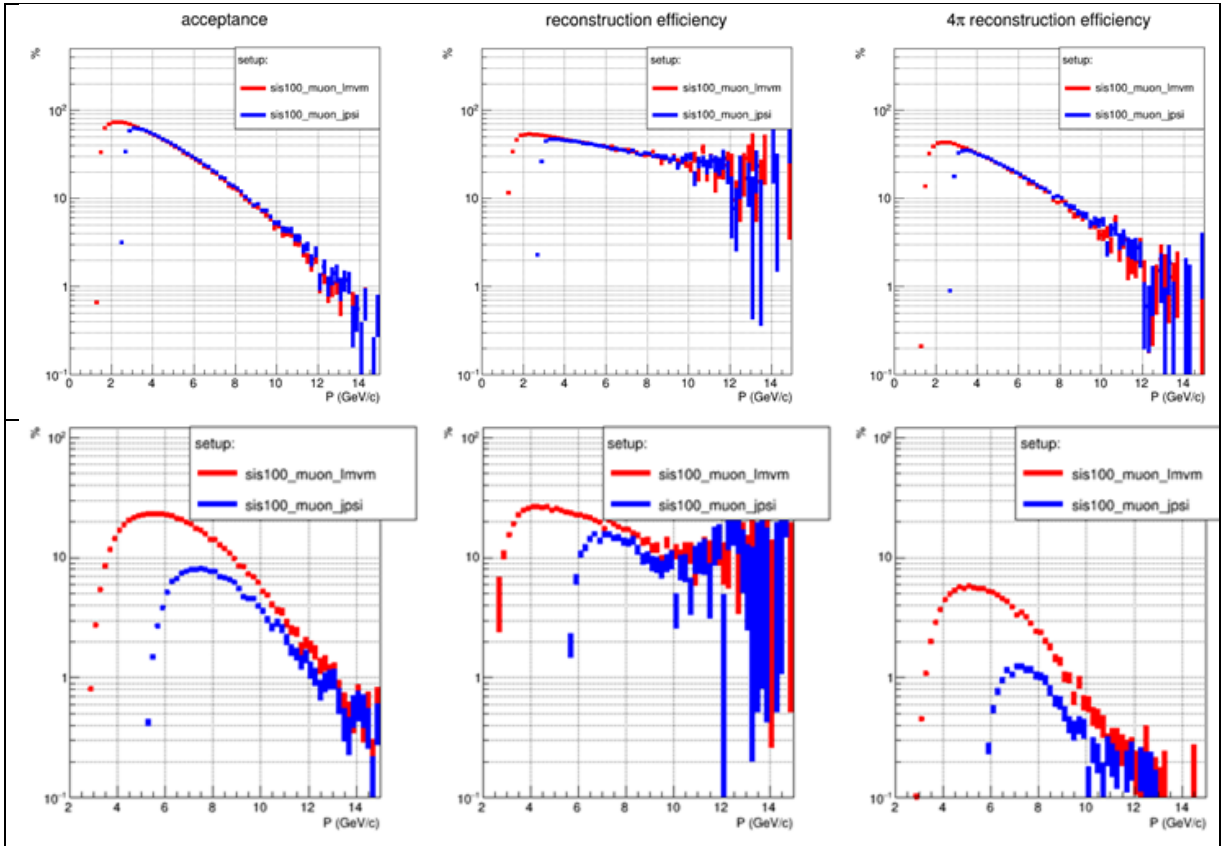

Fig. 5: 4π acceptance (left column), efficiency within the acceptance (center column), and efficiency with respect to 4π (right column) for muons from ω meson decays (ω →$\mu^+\mu^-$, upper row) as function of muon momentum, and for ω mesons decaying in muon pairs as function of the ω momentum (lower row), simulated for central Au+Au collisions at 8A GeV/c. The red symbols correspond to the muon detector configuration with 4 absorbers, which will be used for the measurements of dimuon invariant masses up to 1 GeV/c2. The configuration with 5 absorbers (blue symbols) will be used for the measurement of dimuon invariant masses above 1 GeV/c$^2$ up to the J/ψ, and is shown for comparison.

The acceptance and efficiency values for ω mesons decaying in two muons as depicted in the lower row of figure 5 are calculated by multiplying the corresponding values for the daughter muons according to their momentum. The left lower panel of figure 5 illustrates, that for the setup with 4 absorbers the ω acceptance is well above 10% for ω momenta between 4 and 8 GeV/c, reaching values above 20% in the momentum range from 5 to 7 GeV/c. The lower center panel of figure 5 demonstrates, that for the 4 absorber setup the reconstruction efficiency for ω mesons within the acceptance is well above 10% for the ω momentum range from 3 to 10 GeV/c, and reaches values above 20% for ω momenta between 3.5 and 7 GeV/c. The lower right panel of figure 5 indicates, that the ω reconstruction efficiency with respect to $4\pi$ is above 2% for ω momenta from 3.5 and 8.5 GeV/c, with a maximum of more than 5% around a momentum of 5 GeV/c.

The acceptance for ω mesons in the plane transverse momentum versus rapidity as simulated for central Au+Au collisions at 8A GeV/c is shown in figure 6. The ω mesons are identified by MC information. The left panel of figure 6 depicts the $4\pi$ distribution of ω mesons generated with the PLUTO code. The acceptance for reconstructed ω mesons using the configuration with 4 hadron absorbers is depicted in the center panel of figure 6. The result illustrates, that the hadron absorbers also prevent the reconstruction low-momentum muons from ω decays. For comparison, the right panel of figure 6 indicates the acceptance for reconstructed ω mesons for the setup with 5 hadron absorbers.

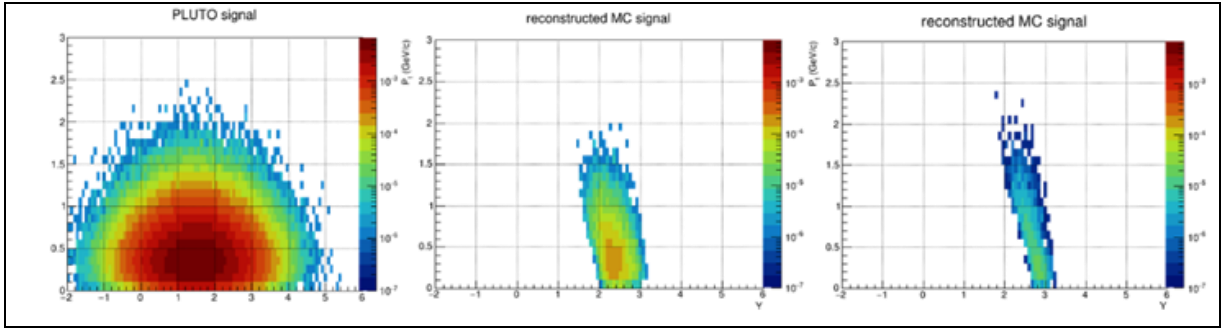

Fig. 6: Transverse momentum as a function of rapidity for ω mesons decaying in muon pairs emitted in central Au+Au collisions at 8A GeV/c. The ω mesons are identified by MC information. Left panel: $4\pi$ distribution as generated with the PLUTO code. Center panel: Acceptance for ω mesons reconstructed after traversing 4 hadron absorbers. Right panel: acceptance for reconstructed ω mesons using the setup with 5 hadron absorbers.

## 4.  Muon measurements and hadron reconstruction

In order to get a complete and consistent picture of the collision dynamics, it is important to correlate different observables event-by-event, such as dileptons and hadrons. For example, if the excitation function of the fireball temperature measured via dileptons indicates a caloric curve, also hadronic observables like event-by-event multiplicity fluctuations of protons should be affected, in order to declare discovery of a first order phase transition in dense QCD matter. Moreover, systematic uncertainties are reduced if different observables are measured simultaneously. In the CBM experiment, electrons and hadrons can be detected in parallel. As will be discussed in the following, also the muon setup allows to identify hadrons, although with limited performance.

### 4.1 Hadron reconstruction with STS without time-of-flight information

Short-lived hadrons can be identified without time-of-flight measurement using only information from the STS. The CBM track reconstruction software based on the Cellular Automaton algorithm [10] provides in addition to the track curvatures also secondary vertices, and, hence, the decay topology of instable particles such as hyperons. The Kalman Filter Particle Finder Package [11] is used to reconstruct the invariant mass of the decay products, using a hypothesis on the particle identities. As examples, the reconstruction results for the decays of $K_S^0 \to \pi^+\pi^-$, $\Lambda \to p\,\pi^-$, $\Xi^- \to \Lambda\,\pi^-$, and $\Omega^- \to \Lambda\,K^-$ are shown in figure 7, simulated for central Au+Au collisions at 12A GeV/c using the UrQMD event generator. $K_S^0$ mesons are reconstructed from 2 tracks of oppositely charged particles emerging from a secondary vertex, assuming that the particles are pions. The same decay topology is used to reconstruct Λ hyperons, assuming that the positive particle is a proton. Once a lambda is identified, which is not created in the primary vertex of the collision, the algorithm looks for another decay product, which is assumed to be either a pion or a kaon. In the first case, $\Xi^-$ hyperons are reconstructed, in the second case $\Omega^-$ hyperons. In both cases the multi-strange hyperons are tracked back to the primary vertex.

As illustrated in figure 7, the reconstruction of $\Omega^-$ hyperons suffers from the large background of pions, which have been considered as kaons. Also the invariant mass spectrum of the $K_S^0$ meson has a rather high background resulting in a signal-to-background (S/B) ratio of about 1. The reason for this relatively low S/B ratio is, that due to its short decay length of $c\tau = 2.67$ cm the $K_S^0$ meson decays right after the target in front of the first STS station, and its reconstruction suffers from a large background of primary pions. Although the decay length of the $\Xi$ hyperon ($c\tau = 4.91$ cm) is shorter than the one of the $\Lambda$ ($c\tau = 7.89$ cm), the background in the $\Xi$ spectrum is lower (S/B=4.6) than for the $\Lambda$ hyperon (S/B= 3.4). The reason is that the $\Xi$ hyperon is reconstructed from three decay particles, which constrains the decay topology better than a two-particle decay. The S/B ratios for both the $\Lambda$ and $\Xi^-$ hyperon are sufficient to extract momentum and angular distributions for further analysis. The total reconstruction efficiency is 22% for $K_S^0$, 23% for $\Lambda$, 9% for $\Xi^-$, and 12% for $\Omega^-$ hyperons.

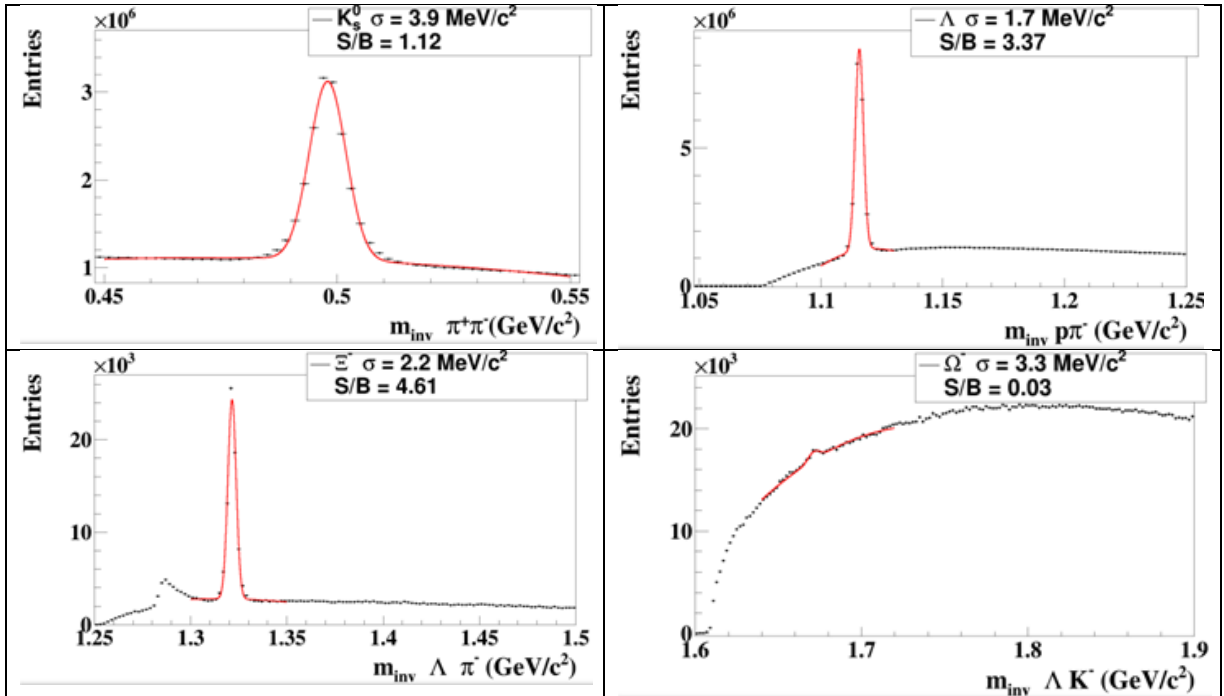

Fig. 7: Invariant mass spectra of $K_S^0$ (upper left panel, $\varepsilon = 22\%$), $\Lambda$ (upper right, $\varepsilon = 23\%$), $\Xi^-$ (lower left, $\varepsilon = 9\%$), $\Omega^-$ hyperons (lower right, $\varepsilon = 9\%$) reconstructed using information from STS only for central Au+Au events at 12 A GeV/c.

In addition to hyperons, also hyper-nuclei like $^3_\Lambda H \rightarrow {}^3He + \pi^-$ or $^4_\Lambda He \rightarrow {}^3He + p + \pi^-$ can be reconstructed using the STS only. In this case, the dE/dx information from the STS is used to separate single charged from double charged particles. The simulated performance of the STS concerning particle identification via dE/dx is illustrated in figure 8 for two cases. Figure 8 depicts the energy loss in a 300 µm thick silicon sensor as function of momentum for particles identified by MC information (left panel), and without particle identification (right panel). In both cases only positive particles are considered, assuming q=1. The red line in the right panel of figure 8 is used to separate particles with charge 1 (dE/dx values below the line) from particles with charge 2 and more (dE/dx values above the line).

Taking into account the dE/dx information from figure 8 (right panel), the invariant mass spectra for the hyper nuclei $^3_\Lambda H$ and $^4_\Lambda He$ have been analyzed in simulations of central Au+Au collisions at 12A GeV/c. In such reactions, $^3_\Lambda H$ and $^4_\Lambda He$ can be reconstructed with signal-to-noise ratios of about 2 and 3.4, and with efficiencies of $\varepsilon = 9.5\%$ and $\varepsilon = 5.7\%$, respectively, as shown in figure 9.

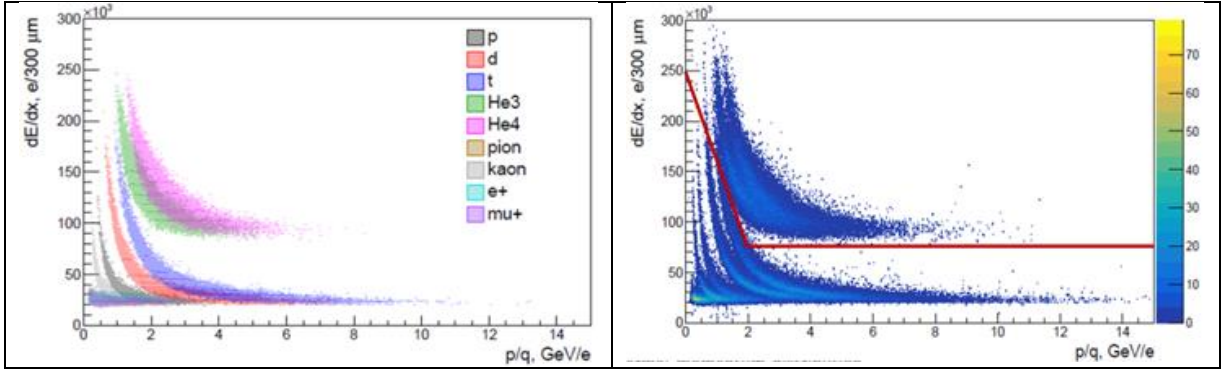

Fig. 8: Calculated energy loss dE/dx in 300 μm silicon as function of particle momentum for single and double positively charged particles assuming q = 1. Left panel: with particle identification from MC. Right panel: without particle identification.

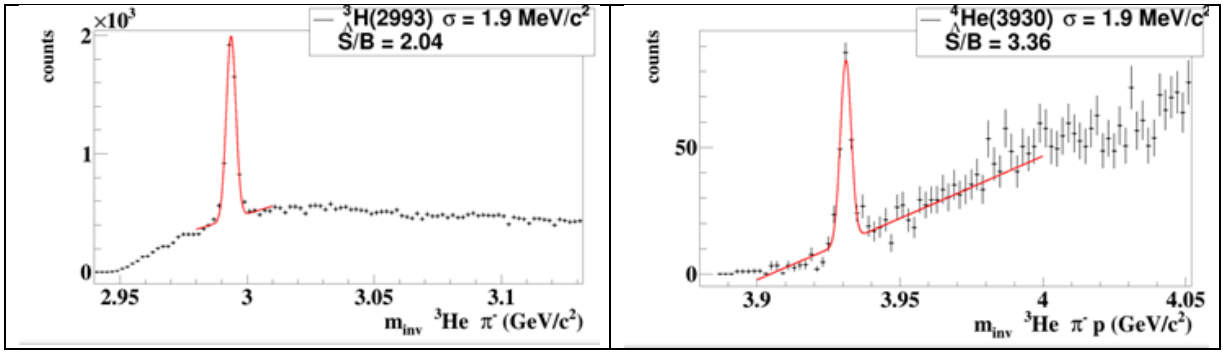

Fig. 9: Invariant mass spectra of $^3$H π$^-$ (left panel) and $^3$H π$^-$ p (right panel), reconstructed in STS using dE/dx information with an efficiency of ε = 9.5% and ε = 5.7%, respectively, in central Au+Au collisions at 12 A GeV/c. The hyper-nuclei $^3_\Lambda$H (left) and $^4_\Lambda$H (right) can be clearly identified with signal-to-background ratios of S/B = 2.04 and S/B = 3.36, respectively.

### 4.2 Hadron identification with TOF behind MUCH

The performance of hadron identification using the CBM setup with 4 hadron absorbers as shown in the central panel of figure 2 is illustrated in figure 10 as a function of the hadron momentum for central Au+Au collisions at 12A GeV/c. The left panel of figure 10 depicts the probability for particles to pass the hadron absorbers and to produce Monte Carlo points in the TRD and in the TOF detector. In this case, only particles are taken into account, which have produced at least 4 Monte Carlo points in the STS, and, hence, can be reconstructed. The kaon yield is reduced by about of a factor of hundred (black dots). Most of the kaons produced at FAIR energies are K$^+$ mesons, which have a small inelastic hadronic cross section, and, therefore, are less suppressed than pions (red dots). On the other hand, the pion yield is 5 to 10 times higher than the kaon yield, depending on the beam energy. Protons are suppressed by a factor of about 1000. The right panel of figure 10 depicts the identification efficiency for the hadrons shown in the left panel, using the TOF information as function of momentum, as illustrated in the right panel of figure 3. Pions with momenta above 2 GeV/C can be identified with an efficiency between 80 and 90%. Similar efficiencies are reached for kaons with momenta above 3 GeV/c. The proton reconstruction efficiency is about 80% for momenta above 3 GeV/c.

The acceptance of the TOF wall for hadrons in the plane transverse momentum versus rapidity during the muon measurements is plotted in figure 11 for kaons (left panel), protons (center panel), and pions (right panel). The simulations were performed for central Au+Au collisions at 11A GeV/c, midrapidity is at y = 1.58. The results demonstrate that the hadron absorbers cut out low transverse momenta around midrapidity and below, but also show, that the remaining range of rapidities is still sufficient to perform reference measurements of hadrons in parallel to the muon experiments.

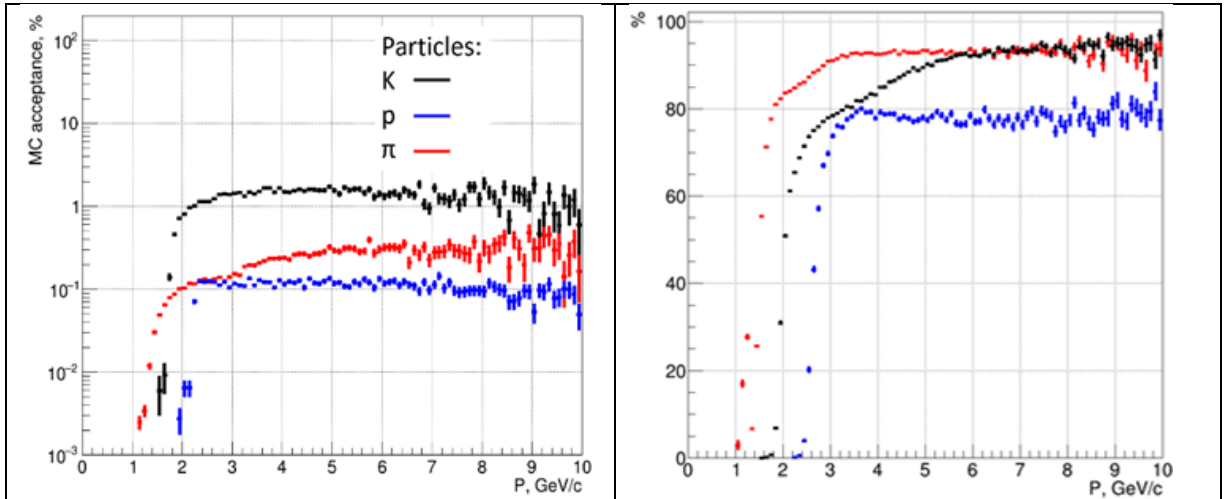

Fig. 10: Left panel: Acceptance of the TOF detector for hadrons traversing the absorbers of the muon detection system as function of the hadron momentum. Right panel: Reconstruction efficiency for hadrons using TOF information during muon measurements (same color code as in the left panel). The simulation was performed fon central Au+Au collisions at 12A GeV/c.

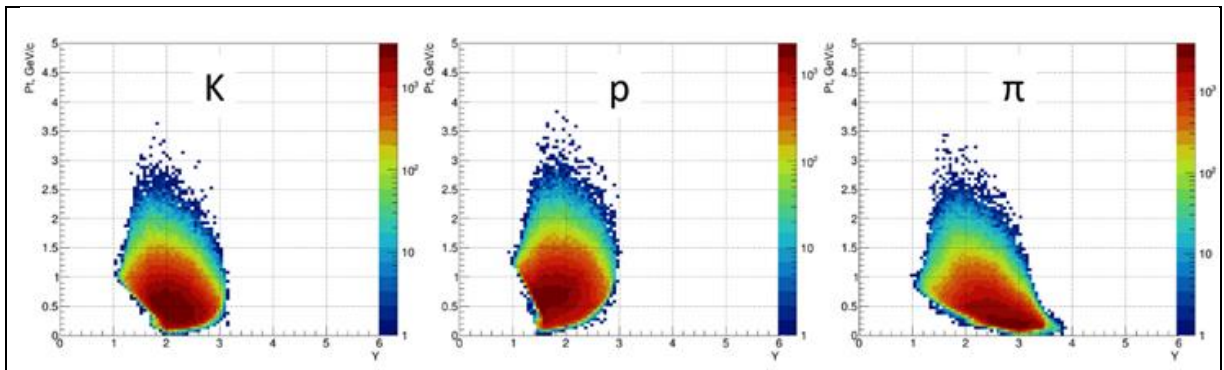

Fig. 11: Acceptance for kaons (left), protons (center) and pions (right) detected in the TOF wall behind the muon system as function of rapidity simulated for Au+Au 11A GeV/c. Midrapidity is at y = 1.58.

## 5. Summary

The Compressed Baryonic Matter (CBM) experiment at the future Facility for Antiproton and Ion Research (FAIR) in Darmstadt is designed to perform multi-differential measurements of hadrons and leptons in central gold-gold collisions at beam energies between 2 and 11 A GeV. The scientific goal of the experiment is to explore the properties of high-density QCD matter. The identification of lepton pairs is notoriously difficult due to the large combinatorial background, which however, is of different origin for electron-positron pairs and muon pairs. Therefore, the measurement of both electrons and muons will substantially reduce the systematic errors of the dilepton data. Moreover, the CBM electron and muon detector systems cover the same experimental acceptance as the hadron detectors, which will provide a consistent and complete picture of the collision obtained from the different observables. For the first time, muon pairs will be measured in heavy-ion collisions at beam energies below the top CERN-SPS energy of 158A GeV. This will be achieved by different configurations of hadron absorbers and tracking detectors, depending on the beam energy. Electrons and muons will be measured alternatively by replacing the respective detector systems. Electrons and hadrons can be identified simultaneously, as the material budget of the RICH detector is low, and hadrons are only moderately rescattered on their way to the Time-of-Flight (TOF) detector located downstream the RICH. The simultaneous measurement of muon pairs and hadrons in the same solid angle, however, poses an experimental challenge. Short-lived particles like K$^0$s and hyperons can be identified by reconstructing their decay topology in the STS only, without identification of their decay products by TOF, although with somewhat increased combinatorial background. The yield of protons, kaons and pions is reduced by the absorbers by two to three orders of magnitude. Nevertheless, the surviving hadrons can be reconstructed and identified via their TOF. Because the muon measurements can be performed with reaction rates of up to 10 MHz, the hyperons, kaons, pions and protons can be recorded simultaneously with very high statistics.

**Acknowledgment**

The muon detection system of the CBM experiment is being developed by a consortium of 13 Indian institutions led by VECC Kolkata and the Bose Institute Kolkata, and by the Petersburg Nuclear Physics Institute in Gatchina. The muon project is supported by the Department of Science and Technology and the Department of Atomic Energy, Govt. of India, and by ROSATOM, Russia. The author acknowledges support from the Europeans Union's Horizon 2020 research and innovation programme under grant agreement No. 871072.